\documentstyle[pra,aps,twocolumn,floats,epsfig]{revtex}

\begin{document}

\wideabs{

\title{Subthermal linewidths in photoassociation
spectra of cold alkaline earth atoms}

\author{Mette Machholm}

\address{Department of Computational Science, The National University of
Singapore, Singapore 119260}

\author{Paul S. Julienne}

\address{National Institute for Standards and
Technology, 100 Bureau Drive, Stop 8423, Gaithersburg, MD 20899-8423}

\author{Kalle-Antti Suominen}

\address{Department of Applied Physics, University of Turku,
FIN-20014 Turun yliopisto, Finland\\
Helsinki Institute of Physics, PL 64, FIN-00014 Helsingin yliopisto, Finland}

\date{\today}

\maketitle

\begin{abstract}
Narrow $s$-wave features with subthermal widths are predicted for the $^1\Pi_g$
photoassociation spectra of cold alkaline earth atoms.  The phenomenon is
explained by numerical and analytical calculations. These show that only a
small subthermal range of collision energies near threshold contributes to the
$s$-wave features that are excited when the atoms are very far apart.  The
resonances survive thermal averaging, and may be detectable for Ca cooled near
the Doppler cooling temperature of the 4$^1$P$\leftarrow$4$^1$S laser cooling
transition.
\end{abstract}
\pacs{34.50.Rk, 34.10.+x, 32.80.Pj}
}
\narrowtext

\section{Introduction}

Photoassociation spectroscopy has become a very powerful tool for studying the
collision physics of laser cooled and trapped atoms~\cite{Weiner99}.  The
conventional wisdom is that the linewidth of individual molecular levels in the
photoassociation spectra of laser cooled atoms is due to the natural linewidth
plus thermal broadening on the order $k_BT$, where $k_B$ is the Boltzmann
constant and $T$ is the temperature.  Thus, we would not expect
photoassociation lines to be much smaller than $k_BT$ in
width~\cite{Napolitano94,Pillet97,Bohn99,Jones99,Burke99,Williams99}.
However, we demonstrate subthermal linewidths for a special case of
photoassociation at very long range to an excited vibrational level $v$ with a
small natural decay width $\Gamma_v$.  In this case the $s$-wave vibrational
features are very narrow at low $T$ (e.g., $<10$~$\mu$K), where $k_BT
\ll\Gamma_v$.  Surprisingly, such features can remain narrow even at much
higher $T$ (e.g., $\approx$ 1~mK), where $k_B T \gg \Gamma_v$.  These
subthermal linewidths are a consequence of having only a narrow range of low
collision energies that contribute to the thermally averaged photoassociation
spectrum.

Narrow features are possible in $^1\Pi_g$ photoassociation trap loss spectra at
small detuning of alkaline earth atoms in a magneto-optic trap
(MOT)~\cite{Machholm01}.  For Ca there is a good chance that the sharp $s$-wave
features will stand out on a background comprised of broader peaks from the
higher partial waves of the $^1\Pi_g$ spectrum and the broad features of the
$^1\Sigma_u$ spectrum, even around the Doppler cooling temperature $T_D$
($T_D=0.83$~mK for the Ca 4$^1$P$\leftarrow$4$^1$S cooling transition). 
Photoassociation spectroscopy in a Ca MOT has been reported~\cite{CaExp}, but
in that experiment the photoassociating laser was detuned far from atomic
resonance (about 780 atomic linewidths $\Gamma_{\rm at}$), whereas our
features are predicted for small detunings ($<$ 25 $\Gamma_{\rm at}$).

\section{Theory of subthermal line shapes}

\subsection{Trap loss spectrum}

In Ref.~\cite{Machholm01} we outlined the numerical and analytical models used
here.  The numerical photoassociation trap loss spectrum is obtained from a
fully quantum mechanical three-channel model, where the time-independent
Schr\"{o}dinger equation is solved with a complex potential to represent
spontaneous decay from the excited state.  We repeat only the essential
analytical formulas here, and concentrate on the explanation of the
phenomenon.  For a more detailed description of different aspects of cold
photoassociation collisions of alkaline earth atoms, see Ref.~\cite{Machholm01}.

\begin{figure}[htb]
\epsfig{width=85mm,file=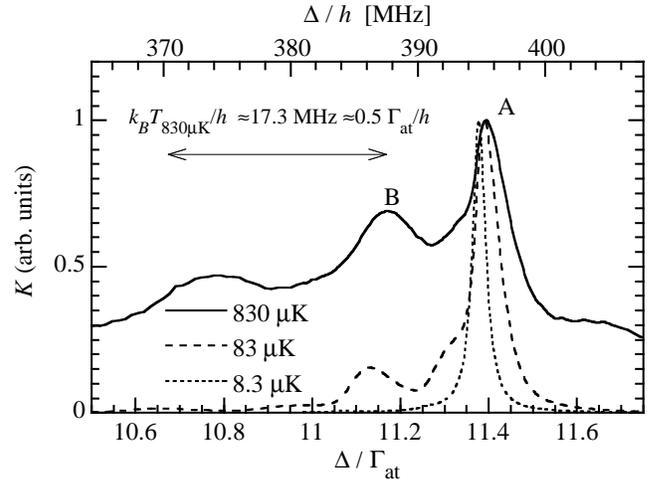} 
\caption[f1]{ Single vibrational feature of the Ca $^1\Pi_g$ photoassociation
spectrum resulting in SC trap loss.  The $\Delta/\Gamma_{\rm at}$ value where
the peak appears depends on the model potentials in Ref.~\cite{Machholm01}.  It
may appear at a different $\Delta/\Gamma_{\rm at}$ in an actual experiment.  The
smooth background of about 2 units due to the $^1\Sigma_g \to {^1}\Sigma_u$
transition~\cite{Machholm01} is not shown.  The Doppler cooling temperature for
cooling on the 4$^1$P atomic state is 830~$\mu$K, and the recoil temperature is
2.7~$\mu$K. The low temperature limit for the feature labeled $A$ is a
Lorentzian line centered at the resonance position $\Delta/\Gamma_{\rm
at}=11.37$ with a natural width of 1.2 MHz.  The peak of the $A$ feature, which
clearly has a subthermal width for $T=830$~$\mu$K, is normalized to unity for
the three different temperatures.  The calculated peak trap loss rate
coefficients for a 1 mW/cm$^2$ laser are $2\times 10^{-13}$~cm$^3$/s (830
$\mu$K), $2\times 10^{-12}$~cm$^3$/s (83 $\mu$K), and $8\times
10^{-12}$~cm$^3$/s (8.3 $\mu$K), respectively.
\label{CaFeature}}
\end{figure}

Figure~\ref{CaFeature} shows a high resolution scan of our calculated
state change (SC) trap loss spectra of cold Ca atoms at three different
temperatures for a single vibrational level of the $^1\Pi_g$ state.  There are
a number of $^1\Pi_g$ isolated vibrational resonance features in the range of
detuning $\Delta$ from 1 to 25 $\Gamma_{\rm at}$ to the red of atomic
resonance.  Our example is typical of these features.  Note that the features
labeled $A$ and $B$ in the figure are both narrower than $k_B T_D$.  Although
the transition between the $^1\Sigma_g$ ground state and the $^1\Pi_g$ excited
state is forbidden at short and intermediate internuclear distances $R$, it
becomes allowed at long range due to relativistic retardation
effects~\cite{Machholm01,Meath68}.  The trap loss collision of the two cold
atoms proceeds via excitation at a long range Condon point $R_C$ to the
$^1\Pi_g$ state (the difference of ground and excited molecular potential
energy curves equals the photon energy at $R_C$).  Once in the $^1\Pi_g$ state
the atoms are accelerated towards short range. The survival probability in
moving from long to short range on the $^1\Pi_g$ state is close to unity due to
the small decay rate ($\Gamma(R)\rightarrow 0$ for $R\rightarrow 0$).  At short
range a SC may occur due to spin-orbit coupling to a lower lying state
correlating to atomic $^1$S + $^1$D, $^3$D or $^3$P states.  After SC to these
channels, modeled here by a single effective channel, the atoms will be lost
from the trap due to the large gain in kinetic energy.

The photoassociation spectrum in Fig.~\ref{CaFeature} is the thermally
averaged loss rate coefficient~\cite{footnote1}:
\begin{eqnarray}\label{thermalrate}
   K(T,\Delta) &=& \sum_{\ell_{even},J}(2\ell+1)
   \frac{\hbar \pi}{\mu} \left \langle
   \frac{|S_{pg}(\varepsilon,\Delta,\ell J)|^2}{k} \right \rangle \nonumber \\
   &=& \sum_{\ell_{even},J} \langle K(\varepsilon,\Delta,\ell J) \rangle .
\end{eqnarray}
Here $\varepsilon = \hbar^2 k^2/(2\mu)$ is the collision energy at momentum
$\hbar k$ for reduced mass $\mu$, $\ell$ is the ground state partial wave
quantum number (0, 2, 4, \dots for identical Group II spinless bosons), $J
=\ell, \ell\pm 1$ is the excited state rotational quantum number, and $\Delta$
is the detuning from the atomic resonance of the photoassociating laser. 
$S_{pg}(\varepsilon,\Delta,\ell J)$ is the $S$-matrix element for the
transition between the ground state $g$ and the SC channel $p$ via the excited
state $e$.  The brackets $\langle \cdot \cdot \cdot \rangle$ imply a thermal
average over a Maxwellian energy distribution:
\begin{equation}
      \langle \cdot \cdot \cdot \rangle = \frac{2}{\sqrt \pi}
      \int_0^\infty x^{1/2} e^{-x} ( \cdot \cdot \cdot ) dx
      \label{maxwellian}
\end{equation}
where $x = \varepsilon / (k_B T)$.  As we show below, a consequence of the
excitation at a very large $R_C$ is that only a small range of energies, much
less than $k_B T$ when $T$ is near $T_D$, contributes to the thermal average
integral in Eq.~(\ref{maxwellian}), especially for $s$-waves.  Consequently,
averaging does not introduce much additional broadening.

\subsection{Analytic theory}

An analytic interpretation can be given for the origin of the subthermal
linewidths.  When the spacing $h\nu_v$ between vibrational levels $v$ is much
larger than their total width $\Gamma_v$, i.e., the vibrational resonances are
non-overlapping, then $|S_{pg}|^2$ is given by an isolated Breit-Wigner
resonance scattering formula for photoassociation
lines~\cite{Napolitano94,Bohn99,Machholm01}:
\begin{eqnarray}
      |S_{pg}(\varepsilon,\Delta,\ell J)|^2 &=& {\Gamma_{vp}
      \Gamma_{vg}(\varepsilon,\ell J) \over \left[ \varepsilon
      - {\cal E}_v(\Delta,J) \right]^2 + \left( {\Gamma_v / 2} \right)^2 }.
      \label{BWres}
\end{eqnarray}
The total width $\Gamma_v$ is the sum of the decay widths into the SC
($\Gamma_{vp}$) and the ground state ($\Gamma_{vg}$) channels and the
radiative decay rate ($\Gamma_{v,rad}$), and ${\cal E}_v (\Delta,J) =
\Delta-[\varepsilon_v(J)+s_v(J)]$ is the detuning-dependent position of the
vibrational level $vJ$ in the molecule-field picture relative to the ground
state separated atom energy.  The level shift $s_v(J)$ due to the laser-induced
coupling~\cite{Bohn99} is small for our case.  When $\Delta= \varepsilon_v
(J)+s_v(J)$, then ${\cal E}_v(\Delta,J)=0$ and the vibrational level is in
exact resonance with colliding atoms with zero kinetic energy.

In the {\it reflection approximation} $\Gamma_{vg}(\varepsilon,\ell J)$ is
proportional to the square of the ground state wavefunction $\phi_g$ at the
Condon point ($R_C$)~\cite{Bohn99,Julienne96,Boisseau00}:
\begin{equation}
     \Gamma_{vg}(\varepsilon,\ell J) = \frac{2\pi h \nu_v |V_{eg}(R_C,\ell
     J)|^2}{D_C} |\phi_g(\varepsilon,\ell,R_C)|^2 .
      \label{RefApprox}
\end{equation}
Here $V_{eg}(R_C,\ell J))$ is the laser-induced coupling, $D_C$ is the slope
difference of the ground and excited state potentials at $R_C$, and $\nu_v$ is
the vibrational frequency for level $v$.  $|V_{eg}|^2$ is linear in laser
intensity $I$ for our assumed weak-field case. Approximating the ground state
wavefunction by its low-energy asymptotic form gives for $s$-waves
\begin{equation}
     |\phi_g(\varepsilon,0,R_C)|^2 =
      \frac{2 \mu}{\pi \hbar^2}\frac{\sin^2 k (R_C-A_0)}{k} ,
      \label{swavePeg}
\end{equation}
where $A_0$ is the scattering length of the ground state potential.
For higher partial waves ($\ell > 0$)~\cite{Taylor},
\begin{equation}
     |\phi_g(\varepsilon,\ell,R_C)|^2 =
      \frac{2 \mu}{\pi \hbar^2} \frac{z_C^2|j_{l}(z_C)|^2}{k} ,
      \label{BesselPeg}
\end{equation}
where $j_{l}(z_C)$ is the spherical Bessel function, and $z_C=k R_C$.
The normal scattering phase shift does not appear in Eq.~(\ref{BesselPeg}) since
it is vanishingly small near threshold for the higher partial
waves~\cite{Weiner99}.  We define here the near threshold range of collision
energies to be the range for which Eqs.~(\ref{swavePeg}) and (\ref{BesselPeg})
are good approximations.

\begin{figure}[htb]
\epsfig{width=85mm,file=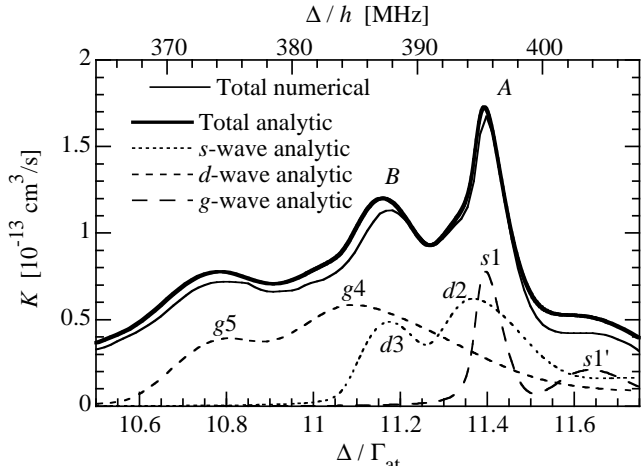}
\caption[f2]{Thermally averaged Ca SC trap loss rate for $T=$ 0.83 mK and $I=$ 1
mW/cm$^2$ from the analytic line shape model,
Eqs.~(\ref{BWres}),~(\ref{swavePeg}) and (\ref{BesselPeg}), compared to the
full quantum numerical result, including contributions from $\ell\leq 6$.  The
separate contributions are shown for the $s$-, $d$-, and $g$-waves, with the
dominant contribution to each peak labeled by $\ell J$.
\label{Analytic}}
\end{figure}

Figure~\ref{Analytic} shows the thermally averaged Ca trap loss spectrum
$K(T,\Delta)$ at $T_D=0.83$ mK obtained by inserting Eq.~(\ref{BWres}) into
Eq.~(\ref{thermalrate}) using Eq.~(\ref{swavePeg}) or (\ref{BesselPeg}).  The
analytic model takes a few input parameters from the numerical
model~\cite{Machholm01}: the scattering length for the ground state potential
(the actual value is unknown, but the model value is 67 a$_0$), $\Gamma_{vp}/h
\approx 0.3$ MHz $< \Gamma_{v,rad}/h = 0.8$ MHz, and the position
$\varepsilon_v(J)+s_v(J)$ of the vibrational levels for each $J$.  The good
agreement with the quantum numerical calculations indicates the quality of the
analytic model.

Figure~\ref{Analytic} also shows the individual contributions from the $s$-,
$d$-, and $g$-waves ($\ell=0,2,4$).  The analytic formulas also compare very
well with the details of these individual features in the numerical calculation
(comparison not shown).  We will concentrate on the overall subthermal features
labeled $A$ and $B$.  The $A$ feature is a sharp $s1$ line sitting on a
background due to $d$ and $g$ lines, whereas the $B$ feature takes its relative
sharpness from a $d3$ line sitting on the $d2$ and $g4$ background.  The $d2$
feature contributes the shoulder on the left of the $A$ peak.

\subsection{Origin of subthermal features}

\begin{figure}[htb]
\epsfig{width=85mm,file=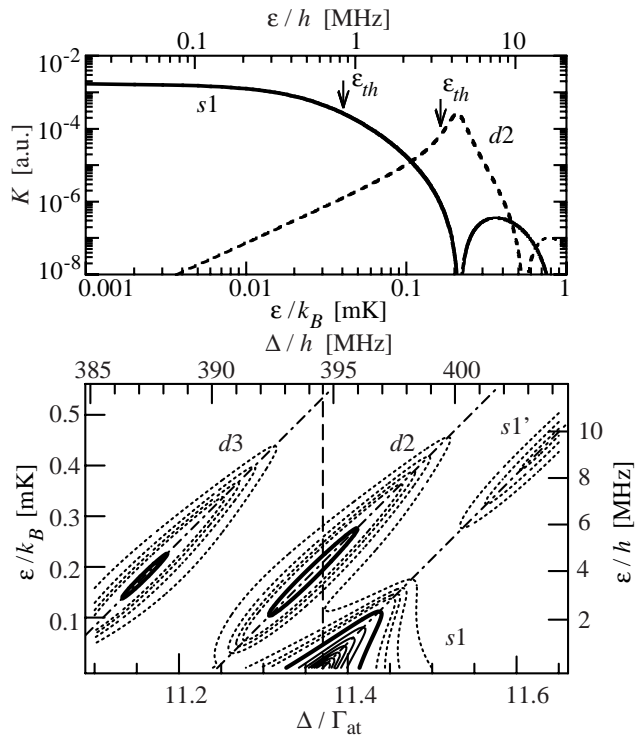}
\caption[f3]{The {\it lower panel} shows the analytic
$K(\varepsilon,\Delta,\ell J)$ in atomic units (1 a.u. $= 6.126 \times
10^{-9}$ cm$^3/$s) versus $\varepsilon$ and $\Delta$ for $\ell J$ = $s1$, $d2$,
and $d3$ ($I=1$ mW/cm$^2$).  The vertical and horizontal axes are set to have
the same energy scale in common units ($k_B/h=$ 20.8 MHz/mK) so that the
dash-dot lines of exact resonance, $\varepsilon={\cal E}_v(\Delta,J)$, have a 45
degree slant.  The dotted contour lines start at 0.000025 and increase in steps
of 0.000025.  The solid contour lines start at 0.0002 (bold) and increase
in steps of 0.0002.  The {\it upper panel} shows a cut of
$K(\varepsilon,\Delta,\ell J)$ for $\ell J =$ $s1$, $d2$ along the dashed
vertical line of constant detuning indicated on the lower panel.  The analytic
threshold range $\varepsilon_{th}$ is indicated.
\label{Smatrix}}
\end{figure}

Figure~\ref{Smatrix} shows $K(\varepsilon,\Delta,\ell J)$ obtained from the
analytic model for the $\ell J =$ $s1$, $d2$, and $d3$ features in
Fig.~\ref{Analytic}.  The slanted maxima follow the lines of exact resonance
where $\varepsilon = {\cal E}_v(\Delta,J)$. Figure~\ref{Smatrix} also shows a
cut of the $s1$ and $d2$ $K(\varepsilon,\Delta,\ell J)$ at the fixed detuning
$\Delta/\Gamma_{\rm at}=11.37$ where ${\cal E}_v(\Delta,J)=0$ for the
$s1$ resonance ($d3$ has a similar energy variation as $d2$, centered
at a different detuning).  The variation of the integrand
$K(\varepsilon,\Delta,\ell J)$ in Eq.~(\ref{maxwellian}) as a function
of $\varepsilon$ and $\Delta$ provides an explanation for the subthermal
features.  Since $K(\varepsilon,\Delta,\ell J)$ peaks in a small range of
$\varepsilon \ll k_BT_D$ , only a small range of collision energies,
$\varepsilon/k_B \approx $ 0.1 mK for the $s$-wave and 0.2 mK for the $d$-wave,
contributes to the width of the feature (for comparison, $\Gamma_{v,rad}/k_B =
0.04$ mK).  Consequently, the $s$-wave peak $A$ broadens only slightly when the
temperature increases from $T$ = 0.083~mK to the Doppler limit $T_D$ = 0.83~mK,
as seen in Fig.~\ref{CaFeature}.  Even the $d$-wave feature $B$ remains
subthermal, although broader than the $s$-wave feature.

The variation of $K(\varepsilon,\Delta,\ell J)$ is a consequence of the
near-threshold resonance form of $S_{pg}(\varepsilon,\Delta,\ell J)$,
Eq.~(\ref{BWres}).  The resonant denominator causes the largest
contribution to $K(\varepsilon,\Delta,\ell J)$ to come from energies near the
slanted line of exact resonance, $\varepsilon = {\cal E}_v(\Delta,J)$,
indicated in the lower panel of Fig.~\ref{Smatrix}.  On the other hand, the
term $\Gamma_{vg}(\varepsilon,\ell J)$ [Eq.~(\ref{RefApprox})], proportional to
$|\phi_g(\varepsilon,\ell,R_C)|^2/k$, is strongly influenced by the
near-threshold properties of $\phi_g(\varepsilon,\ell,R_C)$.  We may
distinguish two regimes: $k \ll k_{th}$ where
$|\phi_g(\varepsilon,\ell,R_C)|^2/k \propto k^{2\ell}$, and $k \gg
k_{th}$, where $|\phi_g(\varepsilon,\ell,R_C)|^2/k$ oscillates with an
amplitude decreasing as $1/k^2 \propto 1/\varepsilon$.  Thus, the
integrand $K(\varepsilon,\Delta,\ell J)$ for $s$-waves approaches a
constant value for $k \ll k_{th}$ and drops off rapidly and oscillates
when $k \gg k_{th}$.  This variation is evident in the upper panel of
Fig.~\ref{Smatrix}.  Using Eqs.~(\ref{swavePeg}) and (\ref{BesselPeg}), we
estimate $k_{th}$ from $k_{th} |R_C-A_0| =\pi/2$ for $s$-waves and $k_{th} R_C
= z_1(\ell)$ for $\ell >0$, where the first maximum in $j_{\ell}(z)$ for
positive argument is at $z=z_1(\ell)$.  Taking $R_C$ = 513 a$_0$ and the
arbitrary model value $A_0=67$ a$_0$ for our case gives $\varepsilon_{th}/k_B =
(\hbar k_{th})^2/(2\mu k_B) =$ 0.05 mK for the $s$-wave and 0.18 mK for the
$d$-wave.  Thus, the large $R_C$ leads to the small value for $k_{th}$
and $\varepsilon_{th}$, and consequently to the subthermal linewidth.
For energies higher than $\varepsilon_{th}$,
$K(\varepsilon,\Delta,\ell J)$ has a node at $\varepsilon$ where
$\phi_g(\varepsilon,\ell,R_C)$ has a node at $R_C$, e.g., at
$\varepsilon/k_B =$ 0.21 mK for the $s$-wave and 0.55 mK for the $d$-wave.  A
second maximum in the ground state wavefunction appears at higher energy, 0.48
mK in the case of the $s$-wave.  This is the origin of the maximum in the red
wing of the $s1$ feature labeled $s1'$ in Figs.~\ref{Analytic} and
\ref{Smatrix}.

\begin{figure}[htb]
\epsfig{width=85mm,file=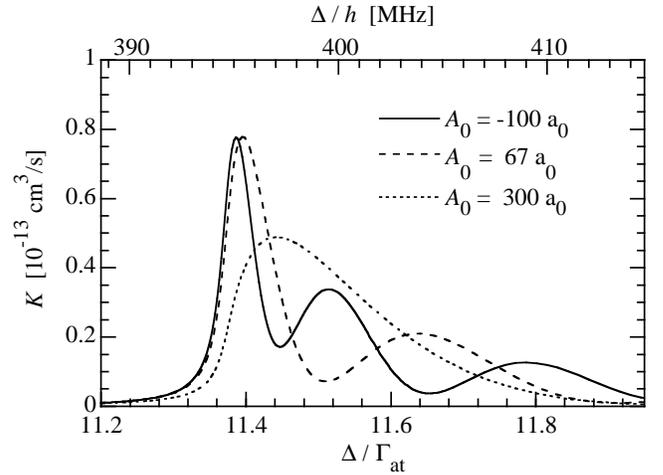} 
\caption[f4]{Variation of the thermally averaged $s1$ feature at 0.83 mK for
three different ground state scattering lengths ($A_0=67$~a$_0$ for
Figs.~\ref{CaFeature}-\ref{Smatrix}).  Thermal averaging causes the peak to
shift from the exact resonance position $\Delta/\Gamma_{\rm at} = 11.37$ (where
${\cal E}_v=0$) depending on the value of $A_0$.  The $s1$ feature broadens and
shifts considerably when $A_0$ becomes close to $R_C$.  In this case,
$\varepsilon_{th}$ increases and a much larger range of collision energies
contribute to the thermal average.  The feature has a linewidth closer to the
natural linewidth of 1.2 MHz and shows more prominent subsidiary maxima
when the scattering length is negative.  In this case, $\varepsilon_{th}$
decreases and a smaller range of energies contribute to the thermal average.
\label{A0depend}}
\end{figure}

Figure~\ref{A0depend} illustrates the dependence of the $s1$ feature on the
unknown ground state scattering length $A_0$.  Since $k_{th} =
\pi/(2|R_C-A_0|)$, $k_{th}$ decreases for $A_0 < 0$, resulting in narrower
thermally averaged lines.  However, if $A_0$ is positive and near $R_C$ so that
$|R_C-A_0|$ becomes small, then $k_{th}$ increases and the narrow peak broadens
and flattens, so that it may no longer stand out.

\section{Conclusion}

We predict that subthermal line shapes should appear in high resolution
photoassociation spectra of the ${^1}\Pi_g$ state of Ca dimer near the
${^1}$P$\leftarrow{^1}$S laser cooling transition. Such features will be hard
to see for Mg, where weak ${^1}\Pi_g$ lines are obscured by a large
${^1}\Sigma_u$ background~\cite{Machholm01}. Subthermal lines may be less
prominent for Sr or Ba because of additional predissociation broadening and a
higher density of states that combine to give broader and more blended
${^1}\Pi_g$ features.

Subthermal linewidth of scattering resonances are possible when the
contributions to the line shape from the relevant $S$-matrix elements is
restricted to very low collision energies below the range of thermal energies
$k_B T$.  In our present study, this is a consequence of the very large Condon
points associated with the transitions.  It would be useful to extend this
analysis to other photoassociative transitions or magnetically-induced Feshbach
resonances~\cite{Vuletic99}.  This would be most interesting in the
case of a large $s$-wave scattering length, that is, when $|A_0| \gg x_0$, where
$x_0=\frac{1}{2}(2 \mu C_6/\hbar^2)^{1/4}$ is a characteristic length scale for
a van der Waals potential with dispersion coefficient $C_6$~\cite{Williams99}. 
Since the near threshold range is small, $0 < k \ll |A_0|^{-1}$, it would be
interesting to see if subthermal lineshapes are possible for very narrow
resonances with widths $\ll k_B T$ when $|A_0|$ is large enough that
$\hbar^2/(2 \mu A_0^2) \ll k_B T$.

\acknowledgments

We thank Nils Andersen and Jan Thomsen of the {\O}rsted Laboratory of the
University of Copenhagen for their hospitality.  This work has been supported
by the Carlsberg Foundation, the Academy of Finland (projects No. 43336 and
No. 50314), the European Union Cold Atoms and Ultraprecise Atomic Clocks
Network, and the US Office of Naval Research.


\begin{references}
\bibitem{footnote1} This equation, together with
Eq.~(\ref{maxwellian}), is equivalent to Eq.~(22) in
Ref.~\cite{Machholm01}, but written in a different form.
\end{references}
\end{document}